# Dissecting spin-phonon equilibration in ferrimagnetic insulators by ultrafast lattice excitation

Sebastian F. Maehrlein[1,2,#], Ilie Radu[3,4,5*], Pablo Maldonado[6], Alexander Paarmann[1], Michael Gensch[7], Alexandra M. Kalashnikova[8], Roman V. Pisarev[8], Martin Wolf[1], Peter M. Oppeneer[6], Joseph Barker[9], Tobias Kampfrath[1,2*]

1. Fritz Haber Institute of the Max Planck Society, Faradayweg 4-6, 14195 Berlin, Germany
2. Department of Physics, Freie Universität Berlin, Arnimallee 14, 14195 Berlin, Germany
3. Max Born Institute for Nonlinear Optics and Short Pulse Spectroscopy, Max-Born-Straße 2A, 12489 Berlin, Germany
4. Helmholtz-Zentrum Berlin für Materialien und Energie, Albert-Einstein-Straße 15, 12489 Berlin, Germany
5. Institute for Optics and Atomic Physics, Technical University Berlin, Hardenbergstraße 36, 10623 Berlin, Germany
6. Department of Physics and Astronomy, Uppsala University, Box 516, 75120 Uppsala, Sweden
7. Helmholtz-Zentrum Dresden-Rossendorf, Bautzner Landstr. 400, 01328 Dresden, Germany
8. Ioffe Institute, 26 Polytechnicheskaya, 194021 St. Petersburg, Russia
9. Institute for Materials Research, Tohoku University, Sendai 980-8577, Japan
\# Present address: Department of Chemistry, Columbia University, 3000 Broadway, New York, NY 10027, USA

\* Corresponding authors. Email: radu@mbi-berlin.de, tobias.kampfrath@fu-berlin.de

**ABSTRACT**

To gain control over magnetic order on ultrafast time scales, a fundamental understanding of the way electron spins interact with the surrounding crystal lattice is required. However, measurement and analysis even of basic collective processes such as spin-phonon equilibration have remained challenging. Here, we directly probe the flow of energy and angular momentum in the model insulating ferrimagnet yttrium iron garnet. Following ultrafast resonant lattice excitation, we observe that magnetic order reduces on distinct time scales of 1 ps and 100 ns. Temperature-dependent measurements, a spin-coupling analysis and simulations show that the two dynamics directly reflect two stages of spin-lattice equilibration. On the 1-ps scale, spins and phonons reach quasi-equilibrium in terms of energy through phonon-induced modulation of the exchange interaction. This mechanism leads to identical demagnetization of the ferrimagnet's two spin-sublattices and a novel ferrimagnetic state of increased temperature yet unchanged total magnetization. Finally, on the much slower, 100-ns scale, the excess of spin angular momentum is released to the crystal lattice, resulting in full equilibrium. Our findings are relevant for all insulating ferrimagnets and indicate that spin manipulation by phonons, including the spin Seebeck effect, can be extended to antiferromagnets and into the terahertz frequency range.

# INTRODUCTION

In solids, vibrations of the crystal lattice have a significant impact on the orbital dynamics of the electrons (Fig. 1A). They strongly modify properties such as electrical conductivity and may even cause insulator-to-metal transitions[1]. Likewise, the interplay between phonons and electron spins (Fig. 1A) is relevant for equally drastic phenomena including colossal magnetoresistance, femtosecond magnetization control[2,3,4,5,6] and the spin Seebeck effect[7,8]. Recently, ultrafast optical techniques have provided new insights into the ultrafast coherent coupling of individual phonon and spin modes[9,10,11,12,13,14].

Despite this progress, the microscopic origins of spin-phonon interactions remain an intriguing problem. Even the equilibration of crystal lattice and electron spins, arguably the conceptually simplest collective process, is far from being understood. This notion is highlighted in the model system yttrium iron garnet $Y_3Fe_5O_{12}$ (YIG), which is one of the best-studied magnetically ordered solids[15,16] and ubiquitous in the field of magnonics[8]. Only estimates of the time constant of spin-phonon equilibration exist, and they extend over as many as 6 orders of magnitude from ~1 µs (Ref. 17) to ~250 ps (Refs. 18 and 19) and down to ~1 ps (Ref. 20). Here, we introduce an approach based on resonant phonon excitation that allows us to directly probe the interactions between the crystal lattice and electron spins over multiple time scales.

**Experiment.** A schematic of our terahertz (THz)-pump magnetooptic-probe experiment is shown in Fig. 1A and detailed in the **Materials and Methods** section and Fig. S1A. An incident, intense, ultrashort THz pump pulse[21] (photon energy of ~0.1 eV, duration of ~250 fs, see Fig. S1B) selectively excites the crystal lattice by resonantly driving infrared-active transverse-optical (TO) phonons. The impact on the sample's magnetic order is monitored from femtoseconds to milliseconds by measuring the magnetooptic Faraday rotation of a time-delayed probe pulse (see Figs. 1A and S1A). In addition, we measure isotropic changes in the optical transmittance of the sample on ultrafast time scales.

As samples, we choose model systems for spin-wave dynamics in magnetic insulators[7,8,16,22,23]: pure YIG and bismuth/gallium-substituted YIG (BiGa:YIG). In these ferrimagnets, magnetic $Fe^{3+}$ ions at a- and d-sites in the unit cell (Fig. 1B) comprise two inequivalent, ferromagnetic sublattices with magnetization $M_a$ and $M_d$, respectively, which couple antiferromagnetically. The 2:3 ratio of a- to d-sites results in a nonzero net magnetization $M_a+M_d$ below the Curie temperature $T_C$. The Faraday rotation $\theta$ of the probe pulse is determined by[24]

$$\theta = a_a M_a + a_d M_d \tag{1}$$

where $a_a$ and $a_d$ are the sublattice magnetooptic coefficients. Compared to YIG, the Faraday effect of BiGa:YIG is enhanced by about one order of magnitude and of opposite sign[24] (see **Materials and Methods** and Fig. S1C).

In both materials, we expect that the analysis of the pump-induced dynamics is simplified due to the following features. First, the sublattice magnetizations $M_a$ and $M_d$ almost entirely arise from the spin rather the orbital angular momentum of the electrons. At room temperature, the g-factor of YIG differs by only 0.13% from that of the free electron[25]. Second, the sizeable electronic band gap (2.85 eV for YIG) implies that the electronic orbital degrees of freedom remain in their ground state, thereby significantly reducing the complexity of the pump-induced processes.

Figure 2A displays the absorptance spectrum of a 15 µm thick BiGa:YIG film from 10 to 45 THz where absorption is known to be due solely to infrared-active phonons[26]. Our *ab initio* calculations show that the pump pulse centered around 21 THz (red spectrum in Fig. 2A) predominantly excites long-wavelength TO(Γ) lattice normal modes characterized by an asymmetric Fe-O stretch vibration (see Figs. 1B, S4 and **Materials and Methods**).

# RESULTS AND DISCUSSION

**Ultrafast spin dynamics.** Figure 2A shows the relative pump-induced change $\Delta\theta/\theta_0 = \theta(t)/\theta_0 - 1$ in the Faraday rotation as a function of the delay $t$ since excitation. Here, $\theta_0 = \theta(-2 \text{ ps})$ refers to the equilibrium case. When pumping off the TO($\Gamma$) phonon resonances, at around 38 THz (blue pump spectrum in Fig. 2A), a relatively small signal is found (blue curve in Fig. 2B). In marked contrast, we witness a response more than one order of magnitude stronger for resonant phonon excitation around 20 THz (red pump spectrum in Fig. 2A): an ultrafast single-exponential drop of the Faraday signal with a time constant as short as $\tau_{\text{fast}} = 1.6$ ps (red curve in Fig. 2B). On much longer time scales, an additional exponential reduction of $\theta$ with a time constant of $\tau_{\text{slow}} = 90$ ns is found (Fig. 2C). Recovery back to the initial state occurs over about 1 ms (Fig. S2D).

Since the pump-probe signal grows linearly with the pump fluence (inset of Fig. 2C), excitation is dominated by one-photon absorption, whereas strong-field effects such as field or impact ionization are negligible[27]. This notion is further supported by Fig. 2A which demonstrates that the sample absorptance and transient Faraday rotation are found to depend on the pump frequency in a very similar way.

We emphasize that almost identical dynamics are observed when a pure YIG sample instead of the BiGa:YIG film is used (see Fig. S2B-D). Additional control experiments confirm that, as soon as the pump pulse has left the sample, the transient Faraday signal $\Delta\theta(t)$ reliably reflects the dynamics of the sublattice magnetizations $M_a$ and $M_d$ with time-independent coefficients $a_a$ and $a_d$ in Eq. (1) (see **Materials and Methods** and Fig. S3).

Figure 2B also shows the relative optical transmittance change of the BiGa:YIG sample for the case of resonant pumping. The signal starts with a peak-like feature whose width coincides with that of the THz pump pulse. Already for delays $t > 1$ ps, the signal is almost constant, changing by less than 10% in the subsequent evolution. Such relaxation to a quasi-steady state is substantially faster than seen for the transient Faraday rotation which still doubles its value at $t = 1$ ps in the following 5 ps (Fig. 2B). Furthermore, the transmittance signal is found to be independent of the sample magnetization (Fig. S3C). These features suggest that the transmittance change predominantly monitors the redistribution dynamics of the pump-deposited energy in the crystal lattice.

Although the relative signal changes are small and still in the perturbative regime, our results in Fig. 2 show a proof of concept that resonant phonon excitation provides an ultrafast manipulation of magnetic order. It only involves the crystal lattice and electron spins, yet no electronic orbital degrees of freedom (Fig. 1A). The picosecond spin dynamics observed here (Fig. 2B) are unexpected because they are five orders of magnitude faster than the spin coherence lifetime (>0.1 µs) of YIG, which is known to be one of the longest amongst magnetically ordered materials[8,15].

**Temperature dependence.** To characterize the transient states established on the fast ($\tau_{\text{fast}} = 1.6$ ps) and slow ($\tau_{\text{slow}} = 90$ ns) time scale, respectively, we increase the sample temperature $T_0$ from 300 K to 420 K. Simultaneously, we measure the equilibrium Faraday rotation $\theta_0$ as well as its pump-induced change $\Delta\theta$ at ultrashort ($t = 10$ ps, Fig. 2B) and long delays ($t = 1$ µs, Fig. 2C) after phonon excitation.

The resulting $\theta_0$ versus $T_0$ has the typical shape[15] of a ferrimagnet's static magnetization curve (Fig. 3A). Its slope $\partial\theta_0/\partial T_0$ steepens with rising $T_0$ until the transition into the paramagnetic phase occurs at the Curie temperature $T_C = 398$ K of BiGa:YIG. In contrast, the pump-induced Faraday signal at $t = 1$ µs (Fig. 3B) increases with $T_0$ and reaches a maximum right below $T_C$, reminiscent of the derivative of the static curve (Fig. 3A). Indeed, we find that $\Delta\theta(1\,\mu\text{s})$ versus $T_0$ closely follows $(\partial\theta_0/\partial T_0)\Delta T$ (Fig. 3B). Here, $\Delta T = 0.39$ K is the increase in equilibrium temperature as calculated from the energy density deposited by the 1 µJ pump pulse (see **Materials and Methods**). The good agreement of both curves shows that ~1 µs after pumping, the BiGa:YIG film is in full thermodynamic equilibrium characterized by temperature $T_0+\Delta T$.

Remarkably, Fig. 3B reveals that the changes in the Faraday signal at 10 ps are systematically yet nonuniformly smaller than at 1 µs, in agreement with Figs. 2B,C. While for temperatures $T_0<380$ K, $\Delta\theta(10\text{ ps})$ amounts to roughly 80% of $\Delta\theta(1\text{ µs})$, this ratio decreases strongly when $T_0$ is increased. For example, slightly above 380 K (where both curves reach their maxima) and slightly below 400 K, $\Delta\theta(10\text{ ps})$ has reduced to only 50% and 20% of $\Delta\theta(1\text{ µs})$, respectively. Consequently, $\Delta\theta(10\text{ ps})$ vs $T_0$ in Fig. 3B cannot be made to agree with $\Delta\theta(1\text{ µs})$ vs $T_0$ by a simple global rescaling of the $\Delta\theta(10\text{ ps})$ values. In particular, the shape of $\Delta\theta(1\text{ µs})$ vs $T_0$ agrees much better with a suitably scaled derivative $\partial\theta_0/\partial T_0$ of the static Faraday rotation (black curve in Fig. 3B) than $\Delta\theta(10\text{ ps})$ vs $T_0$. Therefore, at time $t=10$ ps after pump excitation, the spin system is in a state that is significantly different from the equilibrium state found at $t=1$ µs.

**Analysis of spin couplings.** To understand the microscopic mechanism driving the ultrafast change in magnetic order following resonant phonon excitation (Fig. 2B), we note that solids exhibit only three fundamental spin couplings. They can be understood as effective magnetic fields exerting torques on spins. In the following, we discuss all of them in terms of their capability to modify the sublattice magnetizations $M_a$ and/or $M_d$ on ultrafast time scales.

First, spin-orbit (SO) coupling is often considered to dominate the ultrafast demagnetization of laser-excited ferromagnetic metals in the absence of transport[28]. In our experiment, however, we find identical ultrafast dynamics for pure YIG and BiGa:YIG (see Fig. S2B), even though the Bi ions are known to increase SO coupling. This notion is manifested in the magnetooptic effects which are enhanced by more than one order of magnitude[24]. Therefore, SO coupling plays a negligible role in the ultrafast spin dynamics seen in Fig. 2B.

The same conclusion can be drawn for the second type of coupling, spin-spin magnetic-dipole (SSMD) interaction, which has a comparable strength to SO coupling in YIG[15]. We note that SO and SSMD coupling also account for magnetic anisotropy, which can undergo strong photo-induced changes in the garnet YIG:Co (Ref. 29). Ultrafast laser-induced modification of the anisotropy field of a YIG:Co film was even shown to drive precessional magnetization switching[29]. The observed time constant of this process (~20 ps) is, however, still more than one order of magnitude longer than the ultrafast dynamics ($\tau_{\text{fast}}=1.6$ ps) observed here. In addition, we do not observe any sign of coherent precession which is typical of anisotropy changes in the perturbative regime. Consequently, neither SO nor SSMD coupling can explain the spin dynamics observed on the $\tau_{\text{fast}}$ scale here.

We finally consider the only remaining fundamental spin coupling, isotropic exchange, which is the strongest spin interaction in most magnets[2]. In YIG, it is responsible for the ferrimagnetic order and the high magnon frequencies extending to more than 20 THz at room temperature[15,30]. Therefore, exchange interaction may well account for the picosecond spin dynamics observed here (Fig. 2B). Importantly, since it conserves the total spin angular momentum, the magnetization changes of a- and d-sublattice cancel each other. Despite $\Delta M_a+\Delta M_d=0$, the resulting change in Faraday rotation is nonzero because the magnetooptic coefficients $a_a$ and $a_d$ of YIG differ significantly[24] (see Eq. (1)). In summary, our analysis of the fundamental spin couplings shows that only isotropic exchange interaction is capable of explaining the picosecond spin dynamics seen in our experiment.

**Model.** We suggest the following scenario of exchange-mediated spin-phonon coupling: the resonantly excited TO(Γ) phonons decay into other modes significantly faster than the time constant $\tau_{\text{fast}}=1.6$ ps of the measured ultrafast spin dynamics[26], thereby heating up the crystal lattice. The assumption of ultrafast phonon relaxation is consistent with our transient transmittance data (Fig. 2B) which indicate that the phonons reach an approximately stationary state within less than 1 ps. As the heat capacity of the crystal lattice of YIG is two orders of magnitude higher than that of the spin system[31], the lattice acts akin to a bath whose temperature is suddenly increased from $T_0$ to $T_0+\Delta T$.

As a consequence of the ultrafast lattice heating, the $O^{2-}$ ions, which are the lightest, will undergo additional random deflection $\Delta u(t)$ (Fig. 4A). This perturbation modulates the superexchange[32] of adjacent a-$Fe^{3+}$ and d-$Fe^{3+}$ spins and the associated coupling constant $J_{ad}$ by[16,33,34]

$$\Delta J_{ad}(t)=(\partial J_{ad}/\partial u)\, \Delta u(t). \tag{2}$$

We now put this model to test by conducting calculations of both dynamic and equilibrium states and by their comparison to our experimental observations on the ultrafast ($\tau_{fast}$, Fig. 2B) and the slower time scale ($\tau_{slow}$, Fig. 2C).

**Dynamics on the $\tau_{fast}$ scale.** To determine the ultrafast rate of change of $M_a$ und $M_d$, atomistic spin-dynamics simulations based on ~$10^6$ coupled spin equations of motion are performed[30] (see **Materials and Methods**). We include time-dependent exchange parameters through Eq. (2) where $\Delta u(t)$ is assumed to be random with a variance given by the pump-induced temperature increase $\Delta T$ of the ionic lattice.

Simulation results are shown in Fig. 4B. At times $t<0$, the magnetizations of both sublattices fluctuate around their constant means $M_{a0}$ and $M_{d0}$. However, when fluctuations $\Delta J_{ad}(t)$ of the exchange parameter are switched on at $t=0$, $M_a$ decreases linearly with time, until $\Delta J_{ad}(t)$ is switched off. We find the opposite behavior for $\Delta M_d$. The sum curve $\Delta M_a+\Delta M_d$ has a more than one order of magnitude smaller amplitude. It arises from the much weaker random fields of the bath due to spin couplings other than isotropic exchange (see **Materials and Methods**). Figure 4B demonstrates that thermal modulation of $J_{ad}$ induces demagnetization of the two spin-sublattices by the same amount and, thus, transfers energy into the spin system.

The slope of the simulated $\Delta M_a(t)/M_{d0}$ (Fig. 4B) can directly be compared to the initial slope of the pump-probe signal $\Delta\theta(t)/\theta_0$ (Fig. 2B), which amounts to approximately $-0.1\%$ ps$^{-1}$. Agreement between experiment and theory is obtained by choosing $\partial J_{ad}/\partial u$~10 $J_{ad}$ Å$^{-1}$ (see **Materials and Methods**). This value and the theory of superexchange[32] imply that the overlap integral of the ground-state wavefunctions of adjacent $O^{2-}$ and $Fe^{3+}$ ions changes by 100% when their distance changes by ~0.4 Å. As the distance between the two ions is one order of magnitude smaller than the lattice constant $a$ of YIG[15], we estimate $\partial u/\partial a$~0.1 and obtain $\partial J_{ad}/\partial a$~$(\partial J_{ad}/\partial u)(\partial u/\partial a)$~1 $J_{ad}$ Å$^{-1}$. This result is in excellent agreement with recent *ab initio* calculations[34] of the exchange-coupling constants as a function of $a$, which yielded $\partial J_{ad}/\partial a$~0.7 $J_{ad}$ Å$^{-1}$. Therefore, phonon-modulated exchange coupling successfully and consistently explains the rate $(\partial\Delta\theta/\partial t)/\theta_0$ at which magnetic order of YIG is observed to reduce immediately after resonant phonon excitation.

According to our model, $\partial M_a/\partial t$, $\partial M_d/\partial t$ and, thus, $\partial\Delta\theta/\partial t$ (Eq. (1)) are proportional to the difference between lattice and spin temperature (Eq. (10)), in agreement with the linear pump-fluence dependence of the transient Faraday rotation $\Delta\theta$ seen in the experiment (inset of Fig. 2C). Consequently, the measured exponential decay of $\Delta\theta$ to a constant value (Fig. 2B) indicates that the electron spins approach the lattice temperature $T_0+\Delta T$ with the time constant $\tau_{fast}=1.6$ ps.

Note that the subsequent slower dynamics evolve with a time constant of $\tau_{slow}=90$ ns. We conclude that for all pump-probe delays $t\ll\tau_{slow}$, the transient changes in $M_a$ and $M_d$ arise solely from inter-sublattice exchange coupling. As this interaction conserves the total spin angular momentum, the dynamics are under the constraint $\Delta M_a+\Delta M_d=0$ throughout that time window.

**Dynamics on the $\tau_{slow}$ scale.** As indicated by our temperature-dependent measurements (Fig. 3), the garnet film approaches full equilibrium at temperature $T_0+\Delta T$ with the time constant $\tau_{slow}$. The total magnetization $M_a+M_d$ of this state is lower than for $t \ll \tau_{slow}$ where the change $\Delta M_a+\Delta M_d$ is still zero. Thus, transfer of angular momentum from the spin system to the lattice has certainly occurred for times $t \sim \tau_{slow}$ and above.

The only microscopic spin couplings capable of changing the total spin and, thus, magnetization $M_a+M_d$ are SO and SSMD coupling. If SO coupling were dominant, one would expect a difference in the spin dynamics in pure YIG and in BiGa:YIG in which SO coupling is enhanced by partial Bi substitution. However, the signal-to-noise ratio of the Faraday signal seen on the $\tau_{slow}$ scale of pure YIG (Fig. S2B) does currently not allow us to draw a conclusion about which of the two couplings is dominant.

**Constrained spin state.** As discussed above, the spin system is under the constraint of constant magnetization $M_a+M_d=M_{a0}+M_{d0}$ for times $t \ll \tau_{slow}$ whereas it is in full, unconstrained equilibrium for $t \gg \tau_{slow}$ (Fig. 2C). Assuming that for $t \gg \tau_{fast}$ the spins and crystal lattice are characterized by a single temperature $T_0+\Delta T$, both the constrained and unconstrained spin state can be fully described by equilibrium statistical physics (see **Materials and Methods**).

Using a mean-field approximation, we obtain the change $(\partial M_l/\partial T_0)\Delta T$ in the sublattice magnetizations vs $T_0$ for heating with and without the constraint of conserved $M_a+M_d$ (Fig. 5A). Interestingly, the constrained $\partial M_a/\partial T_0$ of minority spin-sublattice a (green line in Fig. 5A) follows closely its unconstrained counterpart. In contrast, the constrained $\partial M_d/\partial T_0$ of majority spin-sublattice d (blue line in Fig. 5A) is systematically smaller than the unconstrained $\partial M_d/\partial T_0$ because sublattice d can only demagnetize as much as sublattice a ($\Delta M_d=-\Delta M_a$). In particular, the difference between constrained and unconstrained $\partial M_d/\partial T_0$ increases considerably when the temperature $T_0$ approaches the Curie point (Fig. 5A). We emphasize that this behavior is in excellent qualitative agreement with the measured Faraday signals $\Delta\theta(10\ \mathrm{ps})$ and $\Delta\theta(1\ \mathrm{\mu s})$ vs $T_0$ (Fig. 3B).

It is interesting to note that in the equilibrium formalism used here, the constant total magnetization is reinforced by a Lagrange multiplier which can be interpreted as a virtual homogeneous magnetic field (see Eq. (13)). One could also term this field "spin pressure", in analogy to the pressure that is required to keep a gas of particles in a bottle of constant volume while the gas temperature is increased. In our experiment, the "spin pressure" builds up on the time scale given by $\tau_{fast}$ and is released by angular-momentum transfer to the crystal lattice on the time scale $\tau_{slow}$.

**Picture of spin-phonon equilibration.** We summarize that our dynamic and equilibrium calculations are fully consistent with the time scales, fluence dependence, temperature dependence and magnitude of the transient Faraday rotation found in the experiment. This agreement leads us to the following picture of the flow of energy and angular momentum of phonon-pumped YIG (Fig. 5B): [1] the pump pulse excites zone-center TO(Γ) phonons which [2] decay within the pump duration, thereby increasing the crystal-lattice temperature. The additional thermal modulation of a-d-exchange induces [3a] transfer of angular momentum between a- and d-spin-sublattices and implies [3b] energy transfer from the phonon to the spin system with the time constant $\tau_{fast}=1.6$ ps (Fig. 2B). The excess magnetization of this constrained state decays by [4] transfer of angular momentum and energy between crystal lattice and spins with the time constant $\tau_{slow}=90$ ns, resulting in full, unconstrained equilibrium (Fig. 2C). The angular-momentum transfer is mediated by SO and/or SSMD interactions[22] which do not conserve $M_a+M_d$.

# CONCLUSION

We employ a THz-pump magnetooptic-probe experiment to investigate spin-lattice equilibration from the femtosecond to the microsecond time scale. Combined with a spin-coupling analysis and atomistic spin-dynamics simulations, our results reveal that the speed of spin-phonon relaxation in ferrimagnetic insulators critically depends on whether one refers to energy or angular momentum.

On a picosecond scale, spins and phonons reach a quasi-equilibrium through phonon-modulated exchange interaction. It induces redistribution of energy among phonons and spins, whereas angular momentum is only transferred between spins. On a much longer time scale in the nanosecond range, the additional transfer of angular momentum between spins and crystal lattice results in full equilibration. The quasi-equilibrium persists on the intermediate time scale and can be understood as a transient spin state with elevated temperature yet unchanged net magnetization. It could be considered as a realization of magnon populations with nonzero chemical potential which were recently introduced in the theory of magnon transport by the spin Seebeck effect[19].

Transfer of angular momentum between spin-sublattices is also a key process in magnetic switching of ferrimagnetic metallic alloys by optical femtosecond laser pulses[35,36,37,38]. This process is considered to follow the preferential ultrafast demagnetization of one of the two spin-sublattices. Note that the flow of angular momentum can proceed through various mechanisms such as direct exchange torque[36] or nonlocal processes[38] including the transport of spin-polarized electrons. Our results reveal a new mechanism for manipulating the exchange interaction on ultrafast timescales and highlight the important role phonons can play in the direct exchange dynamics between spin-sublattices.

In terms of applications, our results suggest that resonant phonon excitation is a new pathway to the preparation of spin states with increased temperature yet unchanged total magnetization. This route may be particularly interesting for the manipulation of antiferromagnetic order[2], where spin angular momentum is inherently conserved. Finally, the ultrafast spin-lattice coupling of YIG implies that the magnon temperature follows the phonon temperature with a delay on the order of only 1 ps, thereby shifting the cutoff frequency of the bulk spin Seebeck effect[7] to the THz range.

## MATERIALS AND METHODS

**Samples**. Our iron garnet films are grown by liquid-phase epitaxy on substrates of gadolinium gallium garnet $Gd_3Ga_5O_{12}$ (GGG). To prevent pump absorption by the substrate, several iron-garnet films are transferred on diamond windows after mechanically removing the GGG. Test measurements confirm that GGG pump absorption is irrelevant to the ultrafast dynamics.

We study two types of garnet samples: pure yttrium iron garnet $Y_3Fe_5O_{12}$ (YIG) and bismuth/gallium-substituted iron garnet $Bi_xY_{3-x}Ga_yFe_{5-y}O_{60}$ ($x$=1.53, $y$=1.33, BiGa:YIG). They have, respectively, a Curie temperature of 545 K and 398 K, and in-plane and out-of-plane magnetic anisotropy. Film thicknesses cover a range from 7 to 20 µm. Chemical composition of the garnet films is determined by X-ray fluorescence measurements. Substitution of Y by Bi in the BiGa:YIG sample enhances the magnetooptic Faraday effect by one order of magnitude[24]. Typical hysteresis loops obtained by measuring the static Faraday rotation of the probe beam are shown in Fig. S1C.

THz spectra of the sample absorptance (which equals one minus the reflected and transmitted power, both normalized to the incident power) are obtained by combined reflection and transmission measurements with a broadband THz time-domain spectrometer. Figure 2A shows that maximum absorptance occurs at 21 THz, which is 2 THz above the highest-frequency TO($\Gamma$)-phonon resonance[26]. According to Ref. 26, absorption in this frequency range solely arises from TO phonons, while crystal-field and charge-transfer-gap transitions are located at much higher frequencies[24].

**Experimental design.** A THz pump pulse is used to resonantly excite long-wavelength TO phonons of the iron-garnet film under study. The instantaneous magnetic state of the sample is determined by a subsequent optical probe pulse measuring the magnetooptic Faraday effect. In this way, spin dynamics are monitored over a large range of pump-probe delays ranging from femtoseconds to milliseconds. Details of our setup are shown in Fig. S1.

To generate intense THz pump pulses, we employ an amplified Ti:sapphire laser system delivering pulses (energy of 15 mJ, center wavelength of 800 nm, duration of 40 fs, repetition rate of 1 kHz) to drive two optical parametric amplifiers that, in turn, generate intense infrared pulses (pulse energy of 1.5 mJ and 1.2 mJ, center wavelength of 1280 nm and 1400 nm, duration of 50 fs and 50 fs, respectively). By difference-frequency mixing of the two infrared pulses in a nonlinear-optical GaSe crystal[21] (thickness of 1 mm), we obtain intense THz pulses (tunable from 15 to 60 THz, energy of typically 3 to 20 µJ, pulse duration around 250 fs) with stable carrier-envelope phase. For sample excitation, the THz pulse is focused onto the sample surface to a spot with diameter 180 µm. Its transient electric field $E(t)$ is measured by electrooptic sampling (see below). The pulse spectrum is obtained by Fourier transformation of $E(t)$ and shown in Fig. 2A.

Low-noise probe pulses (8 fs, 800 nm, 0.5 nJ) are derived from the pulse train of the Ti:sapphire seed laser (repetition rate of $f_{rep} = 80$ MHz). They traverse the sample collinearly with the pump after a delay $t$. To measure the probe's transient polarization rotation $\Delta\theta(t)$, we employ a polarimeter consisting of a Wollaston-prism polarizer followed by two fast photodiodes. Measurement of the probe ellipticity $\Delta\eta(t)$ is accomplished by putting a quarter-wave plate before the Wollaston prism. The resulting photodiode current is a train of temporally isolated electrical pulses (duration of 5 ns each, pulse-to-pulse distance of $1/f_{rep} = 12.5$ ns), which is sampled using a computer-controlled fast analog-digital converter card[39]. In this way, we are able to measure signals at delays of $t = \tau + j/f_{rep}$ (with integer $j$) between −100 ns and ~1 ms within a single pump shot and with a delay spacing of $1/f_{rep} = 12.5$ ns. The ultrashort offset $\tau$ is set from −2 ps to 200 ps by a variable mechanical delay stage.

Pump-probe signal traces $\Delta\theta_\pm(t)$ are taken for the sample magnetization saturated in opposite directions using an external magnetic field $\boldsymbol{B}_{ext}$ of ±15 mT, respectively. To measure samples with in-plane or out-of-plane anisotropy, the magnetic field is oriented 45° with respect to the incident pump and probe beams. To vary the sample temperature, the sample is mounted on a resistive heater. The sample temperature is measured with two thermocouples and an imaging infrared thermometer, all of them yielding consistent temperature values. To measure the THz electric field, we substitute the sample by a GaSe crystal and measure the transient ellipticity $\Delta\eta(t) \propto E(t)$ induced by the electrooptic effect[21].

Optionally, our setup also permits measurement of the transient isotropic transmittance of the sample. For this purpose, we remove all polarization-sensitive components such as the Wollaston prism and wave plate and detect the probe-pulse energy with one of the two fast photodiodes of the polarimeter. Based on the small attenuation length[26] (~1 µm) of the resonant pump pulse into the iron-garnet sample, we estimate that the temporal blurring of the pump-probe signals due to pump-probe velocity mismatch is smaller than 10 fs and, therefore, negligible.

**Signal analysis.** Up to first order in magnetization, the Faraday rotation $\theta$ (and likewise the ellipticity $\eta$) of the probe's outgoing polarization is a weighted average[24] of the magnetization of the two spin-sublattices $l = a, d$, that is,

$$\theta = \sum \langle a_l M_l \rangle d + b. \qquad (3)$$

Here, $a_l$ are the local magnetooptic coupling coefficients (Verdet constants) of YIG, $M_l = \boldsymbol{e}_{pr} \cdot \boldsymbol{M}_l$ is the magnetization of sublattice $l$ projected on the propagation-direction unit vector $\boldsymbol{e}_{pr}$ of the probe, $d$ is the sample thickness, and $b$ is an offset arising from any magnetization-independent optical anisotropy. The angular brackets $\langle . \rangle$ denote averaging over the probed sample volume.

In equilibrium, the $a_l$ are constant throughout the probed magnetic volume, $\langle a_l M_l \rangle = a_l \langle M_l \rangle$. The impact of the pump pulse makes all quantities in Eq. (3) time-dependent, resulting in relatively small changes $\theta(t) - \theta_0$ of the signal $\theta_0 = \theta(-2\,\mathrm{ps})$ measured 2 ps before arrival of the pump pulse. According to Eq. (3), the nonmagnetic offset $b$ cancels by calculating the asymmetric part

$$\Delta\theta(t) = \frac{\Delta\theta_+ - \Delta\theta_-}{2} \approx \sum \left[ a_{l0} \langle \Delta M_l \rangle + M_{l0} \langle \Delta a_l \rangle \right]. \qquad (4)$$

The first term on the right-hand side scales with a weighted average of the sublattice magnetizations along $\boldsymbol{e}_{pr}$, which is the quantity we are interested in. The second term, however, makes a contribution independent of the $\Delta M_l$ and arises from a possible pump-induced modulation $\Delta a_l$ of the Verdet constants. Figure S3 shows that the second term is negligible and that $\Delta\theta$ reflects the true magnetization dynamics, $\Delta\theta = \sum a_{l0} \langle \Delta M_l \rangle$, once the pump pulse has left the sample.

**Excitation profile and heating.** Since the absorption length of the THz pump pulse[26] (~1 µm) is smaller than the sample thickness (~10 µm), pump excitation is inhomogeneous along the $z$-axis normal to the film plane. Because the pump-induced signal $\Delta\theta(t)$ depends linearly on the absorbed pump fluence $F_{abs}$ (inset of Fig. 2C), the pump-induced change in the local Faraday rotation $\Delta\theta(z,t)$ is proportional to the pump energy density $w_{abs}(z)$ absorbed in the vicinity of the plane at $z$. In other words, one has $F_{abs} = \int_0^d dz\, w_{abs}(z)$ and $\Delta\theta(z,t) = C(t) w_{abs}(z)$ where $C(t)$ captures the temporal dynamics. As a consequence, the total pump-induced signal

$$\Delta\theta(t) = \langle\Delta\theta(z,t)\rangle d = \int_0^d dz\, \Delta\theta(z,t) = C(t) F_{abs}. \tag{5}$$

is independent of the shape of the absorption profile $w_{abs}(z)$. Therefore, without loss of generality, we can consider the absorption profile to be homogeneous with $w_{abs} = F_{abs}/d$. Our argumentation is valid as long as transport of heat from the YIG film into the substrate is negligible. Figure S2D shows this assumption is fulfilled for pump-probe delays up to at least 1 µs.

Assuming the sample has reached (local) thermal equilibrium at $t = 1\,\mu s$, we estimate the average temperature increase $\Delta T(1\,\mu s)$ of the BiGa:YIG film by dividing $F_{abs}$ by the pumped sample volume $[d\times(\text{pump diameter})^2\times\pi/4]$, by the mass density and by the specific heat capacity[31] of YIG. We obtain a value of $\Delta T(1\,\mu s) = 0.39$ K for an incident pump energy of 1 µJ, which can be interpreted as the temperature increase of a homogeneously excited YIG film.

We checked effects due to accumulative heating of the sample by reducing the repetition rate of the pump pulses from 1000 Hz to 500 Hz by means of a mechanical chopper. The magnitude of pump-induced signal that persists after 1 ms is found to be smaller than 10% of the pump-induced signal at $t \sim 1\,\mu s$. It corresponds to a static homogeneous temperature increase of the probing volume of only ~39 mK, which has a negligible influence on our results, including the temperature dependence seen in Fig. 3B.

**YIG *ab initio* calculations.** Our *ab initio* phonon calculations are performed using the finite-displacement method implemented in the open-source package phonopy[40]. The electronic-structure calculation and equilibrium geometry search of YIG are performed within the Density-Functional Theory (DFT) framework, using the Vienna *Ab-initio* Simulation Package[41] (VASP) and adopting the generalized gradient approximation[42] of the exchange-correlation functional. Our calculations and resulting phonon dispersion relations are detailed in the **Supplementary Material**.

From the element-resolved vibrational density of states (Fig. S4), we find that a continuum of modes from ~7 to 20 THz contributes to the displacement of the $O^{2-}$ ion with approximately frequency-independent weight. This result implies that the vibration of the $O^{2-}$ ion has a mean frequency of $\Omega_O/2\pi \sim 14$ THz and a correlation time of $\tau_O \sim 10$ fs.

**Dynamic spin model.** Our consideration of all possible spin-coupling mechanisms (SO, SSMD, exchange, see main text) leads us to the view that the ultrafast regime of the spin dynamics of YIG following phonon excitation arises from isotropic exchange interaction. More precisely, the thermal vibrations of the ionic lattice modulate the a-d-exchange interaction (Fig. 4A), thereby transferring spin angular momentum from the a- to d-spin-sublattice and quenching each sublattice magnetization at the same rate.

To test this hypothesis, we calculate both the rate of spin-angular-momentum transfer between sublattices (based on numerical simulations) and the temperature dependence of the sublattice magnetizations (based on an analytical treatment). Starting point is the Heisenberg spin Hamiltonian of YIG[15],

$$\hat{H} = -\frac{1}{2}\sum_{ll'jj'} J_{ll'}^{jj'} \hat{S}_l^j \cdot \hat{S}_{l'}^{j'} - g\mu_B \sum_{lj} \mathbf{B}_{ext} \cdot \hat{S}_l^j, \qquad (6)$$

where $\hbar \hat{S}_l^j$ is the spin-angular-momentum operator of the total electron spin of an $Fe^{3+}$ ion (spin quantum number $S=5/2$) located at site $j$ in spin-sublattice $l$. Orbital contributions to the magnetic moments are negligibly small[25]. In YIG, there are two different crystallographic sites $l = a, d$ for the $Fe^{3+}$ ions which are, respectively, surrounded by six $O^{2-}$ ions (octahedral sites) and four $O^{2-}$ ions (tetrahedral sites). The exchange constants have been determined to be as follows[15]: if for given $l$, $l'$, the $j$ and $j'$ represent next neighbors, then $J_{ll'}^{jj'} = J_{ll'}$. Otherwise, $J_{ll'}^{jj'} = 0$. Thus, the exchange part of the Hamiltonian is fully determined by the three constants $J_{aa}$, $J_{dd}$ and $J_{ad} = J_{da}$.

Following Eq. (2), the phonon-induced modulation of the exchange interaction is modeled as variation

$$\Delta J_{ad}(t) = \partial_u J_{ad} \, \Delta u(t) \qquad (7)$$

of the exchange coupling constant between $l = a$- and $l' = d$-spin-sublattices. Here, $\partial_u = \partial/\partial u$, and $\Delta u$ is the pump-heat-induced deflection of the $O^{2-}$ ion mediating the superexchange of adjacent a- and d-type $Fe^{3+}$ ions, as depicted in Fig. 4A. This assumption is justified because $J_{ad}$ predominantly involves the a-$Fe^{3+}$ and d-$Fe^{3+}$ ions and the $O^{2-}$ ion in between. Since the $O^{2-}$ ion is the by far lightest ion in the YIG unit cell, its motion is expected to modulate the a-d superexchange most strongly.

**Atomistic spin-dynamics simulations.** To calculate the rate of spin-angular-momentum transfer between sublattices, we conduct numerical simulations based on Eqs. (6) and (7). For the implementation, we assume an ensemble of classical spins and replace each spin operator $\hat{S}_i$ by $s_i\sqrt{(S+1)S}$. Here, $s_i$ is a vector of length one, and the compound index $(lj)$ is summarized in one index $i$. Likewise, the Hamilton operator $\hat{H}$ turns into the Hamilton function $H$. For the exchange constants $J_{aa}$, $J_{dd}$ and $J_{ad}$, we use the values 0.33 meV, 1.15 meV and –3.43 meV, respectively[15].

By applying Langevin theory, a stochastic Landau-Lifshitz-Gilbert-type equation is derived for each spin[30,43,44],

$$\partial_t s_i = -\frac{|\gamma_i|}{(1+\lambda_i^2)\mu_i}\left[s_i \times \mathbf{B}_{eff,i} + \lambda_i s_i \times (s_i \times \mathbf{B}_{eff,i})\right]. \qquad (8)$$

where $\gamma_i$ is the gyromagnetic constant of a spin with magnetic moment $\mu_i s_i$. Damping due to coupling to non-spin degrees of freedom (the so-called bath) is quantified by $\lambda_i = 2\times 10^{-5}$. The effective field $\mathbf{B}_{eff,i}$ acting on $s_i$ has three contributions: a deterministic part due to the exchange fields $\boldsymbol{B}^0_{\text{exch},i} = -\partial H^0 / \partial \boldsymbol{s}_i$ of the adjacent spins for temporally constant $J^0_{ij}$ and two stochastic components $\boldsymbol{\xi}_i(t)$ and $\Delta \boldsymbol{B}_{\text{exch},i}(t)$ that arise from coupling of the spins to the bath. The first field $\boldsymbol{\xi}_i(t)$ transfers spin angular momentum and energy between the spin system and the bath having temperature $T_0$. According to the fluctuation-dissipation theorem, in equilibrium, the fluctuations $\boldsymbol{\xi}_i(t)$ are balanced by a friction force, the second, Gilbert-type term in Eq. (8), thereby driving the spin system into an equilibrium state with temperature $T_0$ (Refs. 30, 43, 44). The time constant of this relaxation process is set by the damping parameters $\lambda_i$ and found to be irrelevant on the picosecond scale considered here.

The second stochastic field $\Delta \boldsymbol{B}_{\text{exch},i} \propto \sum_j \Delta J_{ij} \boldsymbol{s}_j$ arises from the phonon-induced modulation of the exchange interaction (Eq. (7)). It is distinctly different from $\boldsymbol{\xi}_i(t)$ as it causes energy transfer into the spin system, but leaves the total spin angular momentum unchanged. Since the noise field at temperature $T_0$ is already accounted for by $\boldsymbol{\xi}_i(t)$, the variance of $\Delta J_{ij}(t)$ scales with the temperature increase $\Delta T$ of the crystal lattice induced by the pump pulse. More precisely, owing to the equipartition theorem, the pump-induced deflection $\Delta u$ of the $O^{2-}$ ion obeys $m_O \Omega_O^2 \Delta \langle u^2 \rangle = m_O \Omega_O^2 \langle \Delta u^2 \rangle = k_B \Delta T$, where $\Omega_O / 2\pi \sim 14 \, \text{THz}$ is the mean $O^{2-}$-ion vibration frequency (see above).

To determine the initial rate of phonon-driven angular-momentum transfer between the spin-sublattices, phonons and spins are assumed to be thermalized at temperatures $T_0 + \Delta T$ and $T_0$, respectively (see main text). Following our *ab initio* results (see above), $\Delta u(t)$ is assumed to have a correlation function with a width of $\tau_O \sim 10 \, \text{fs}$, which is accordingly modeled as

$$\langle \Delta u(t) \Delta u(t') \rangle = \frac{k_B \Delta T}{m_O \Omega_O^2} \tau_O \delta(t - t'). \tag{9}$$

The equation of motion (Eq. (8)) is integrated numerically for $N = 64^3 = 2.6 \times 10^5$ unit cells at equilibrium temperature $T_0 = 300 \, \text{K}$. At each time step $t$, we calculate the two sublattice magnetizations $M_l = \sum_j s_l^j \sqrt{(S+1)S}/N$ with $l = a, d$. In our numerical implementation, the pump-induced thermal noise of the exchange constants (Eqs. (7) and (9)) can be switched on or off over arbitrary time intervals and with variable strength $(\partial_u J_{ad})^2 \Delta T$.

Results of our atomistic simulations are shown in Figs. 4B and S5. We find the demagnetization rate of each sublattice scales according to

$$\frac{|\partial_t M_l / M_{d0}|}{\Delta T} = (\partial_u J_{ad})^2 f \tag{10}$$

with constant $f = 2 \times 10^{16} \, \text{m}^2 \, \text{J}^{-2} \, \text{K}^{-1} \, \text{ps}^{-1}$. On the other hand, our experiment shows that the Faraday signal $\Delta \theta$ and, thus, $|\partial_t M_l / M_{l0}|$ scale with the pump fluence (see inset of Fig. 2C) and, thus, the initial $\Delta T$ as well. In other words, we have

$$\frac{|\partial_t M_l / M_{l0}|}{\Delta T} \sim \frac{|\partial_t \Delta\theta / \theta_0|}{\Delta T} = g \tag{11}$$

where $g = 1.9 \times 10^{-3}$ K$^{-1}$ ps$^{-1}$ has been determined from the initial slope of $\Delta\theta/\theta_0 \sim \Delta M_l / M_{l0}$ (Fig. 2B) seen in our experimental results. Comparison of Eqs. (10) and (11) yields the estimate $|\partial_u J_{ad}| \sim \sqrt{g/f} \sim 10 J_{ad}$ Å$^{-1}$.

According to the theory of superexchange[32], $J_{ad}$ is proportional to the fourth power of the overlap integral of the $Fe^{3+}$ and the $O^{2-}$ ion. Therefore, upon changing the distance of the two ions, the overlap integral undergoes relative changes on the order of 2.5 Å$^{-1}$.

**Mean-field sublattice magnetization.** Our previous results strongly indicate that spins and phonons reach quasi-equilibrium for times $t \gg \tau_{fast}$ following phonon excitation. As exchange interaction leaves the total spin angular momentum of the electrons unchanged, this thermal state is constrained by the boundary condition of conserved total spin angular momentum. Such constraints are ubiquitous in thermostatics (such as for a gas contained in a closed bottle) and can be treated by equilibrium statistical physics[45].

Our goal is to determine the magnetization of the d- and a-spin-sublattice of YIG in thermal equilibrium for two cases: the unconstrained (UC) situation without boundary condition and the constrained (C) situation with the boundary condition of constant total spin angular momentum,

$$\sum_{lj} \langle \hat{\boldsymbol{S}}_l^j \rangle = \text{constant}. \tag{12}$$

In both cases, we need to calculate the statistical operator $\hat{D}$ of the system, from which the expectation value $\langle \hat{A} \rangle = \text{Tr}(\hat{D}\hat{A})$ of an observable $\hat{A}$ (such as a sublattice magnetization) can be derived. In thermal equilibrium and for both the UC and C case, the statistical operator is given by[45]

$$\hat{D} = \frac{1}{Z} \exp\left( -\beta \hat{H} - \beta \boldsymbol{p} \cdot \sum_{lj} \hat{\boldsymbol{S}}_l^j \right) \tag{13}$$

where $\beta = 1/k_B T$ with $k_B$ being the Boltzmann constant. The partition function $Z$ is determined by the normalization condition $\text{Tr}\,\hat{D} = 1$. In the UC case, we simply set $\boldsymbol{p} = 0$ in Eq. (13), thereby resulting in the standard canonical statistical operator[45]. In the C case, the constraint of conserved spin angular momentum (Eq. (12)) has to be fulfilled by proper choice of the Lagrange multiplier $\boldsymbol{p}$ in Eq. (13). Interestingly, $\boldsymbol{p}$ can be interpreted as a virtual homogeneous magnetic field that is adjusted such to reinforce the boundary condition of Eq. (12). It is analogous to the chemical potential that keeps the number of particles in a given macroscopic system constant.

The general Heisenberg spin Hamiltonian $\hat{H}$ of Eq. (6) is too complex for an analytical calculation of $\hat{D}$. Consequently, we apply the mean-field approximation[32], $\hat{\boldsymbol{S}}_l^j \cdot \hat{\boldsymbol{S}}_{l'}^{j'} \approx \langle \hat{\boldsymbol{S}}_l^j \rangle \cdot \hat{\boldsymbol{S}}_{l'}^{j'} + \hat{\boldsymbol{S}}_l^j \cdot \langle \hat{\boldsymbol{S}}_{l'}^{j'} \rangle - \langle \hat{\boldsymbol{S}}_l^j \rangle \cdot \langle \hat{\boldsymbol{S}}_{l'}^{j'} \rangle$, omit the number $\langle \hat{\boldsymbol{S}}_l^j \rangle \cdot \langle \hat{\boldsymbol{S}}_{l'}^{j'} \rangle$, and allow for alignment exclusively along the $z$-axis unit vector $\boldsymbol{e}_z$, $\langle \hat{\boldsymbol{S}}_l^j \rangle = \langle \hat{S}_{l,z}^j \rangle \boldsymbol{e}_z = \langle \hat{S}_l^j \rangle \boldsymbol{e}_z$, as set by the external magnetic field $\boldsymbol{B}_{ext} = B_{ext} \boldsymbol{e}_z$. In addition, we assume the spins order homogeneously on each sublattice $l$, that is, $\langle \hat{S}_l^j \rangle = \langle \hat{S}_l^0 \rangle$ (sublattice approximation[32]). By evaluating the $J_{ll'}^{jj'}$ of YIG (see above), we finally obtain the mean-field (MF) Hamiltonian

$$\hat{H}_{MF} = -\frac{1}{2}\sum_{ll'} J_{ll'} N_{ll'} \langle \hat{S}_{l'}^0 \rangle \sum_j \hat{S}_l^j - g\mu_B \sum_{lj} B_{ext} \hat{S}_l^j \qquad (14)$$

where the $J_{ll'}$ are defined as above, and $N_{ll'}$ denotes the number of the next neighbors a site in sublattice $l$ possesses in sublattice $l'$. For YIG, one has $N_{aa} = 8$, $N_{dd} = 4$, $N_{ad} = 6$ and $N_{da} = 4$ (Ref. 15). We can now write the statistical operator of Eq. (13) in a very compact form as

$$\hat{D}_{MF} = \frac{1}{Z} \exp\left(-\beta \sum_l F_l \sum_j \hat{S}_l^j\right) \qquad (15)$$

where the effective field

$$F_l = \frac{1}{2}\sum_{l'} J_{ll'} N_{ll'} \langle \hat{S}_{l'}^0 \rangle + g\mu_B B_{ext} + p \qquad (16)$$

captures the mean field (first and second term) and the scalar virtual field $p = \boldsymbol{p} \cdot \boldsymbol{e}_z$ imposed by the constraint of Eq. (12). By taking advantage of the fact that all spin operators in Eq. (15) commutate with each other, we obtain the expectation value

$$\langle \hat{S}_l^0 \rangle = -\frac{1}{\beta}\frac{\partial}{\partial F_l} \mathrm{Tr}\exp\left(-\beta F_l \hat{S}_l^0\right). \qquad (17)$$

for any $Fe^{3+}$ spin of sublattice $l$. Evaluation of this equation leads to an implicit relationship for $\langle \hat{S}_l^0 \rangle$,

$$\langle \hat{S}_l^0 \rangle = S\mathcal{B}_S(\beta S F_l), \qquad (18)$$

where $S = 5/2$ and $\mathcal{B}_S$ is the Brillouin function (Ref. 32). The magnetization $M_l$ of sublattice $l$ is proportional to $v_l \langle \hat{S}_l^0 \rangle$ where $v_l$ is the number of $l$-$Fe^{3+}$ ions per unit cell with $v_d = 12$ and $v_a = 8$.

For the UC case ($p = 0$), Eq. (18) is a system of two coupled equations and solved numerically, thereby yielding the unconstrained $\langle \hat{S}_l^0 \rangle_{UC}$ for each temperature $T_0$. For the C case, the boundary condition of Eq. (12), rewritten as

$$v_d \langle \hat{S}_d^0 \rangle_C + v_a \langle \hat{S}_a^0 \rangle_C = \mathrm{constant}, \qquad (19)$$

adds a third equation to Eq. (18) which determines $p$. Note that in our experiment, we start from an UC equilibrium given by the $\langle \hat{S}_l^0 \rangle_{UC}$ at temperature $T_0$. Subsequently, the pump pulse increases $T_0$ by a small amount $\Delta T$ while keeping the total spin angular momentum constant. Since we are only interested in small relative changes of the $\langle \hat{S}_l^0 \rangle$, we linearize Eqs. (18) and (19) in terms of $\Delta T$ and solve analytically for the small changes $\Delta \langle \hat{S}_l^0 \rangle_C$.

To be consistent with our atomistic spin-dynamics model (see above), we choose the ratio of the coupling parameters $J_{ad}$, $J_{aa}$ and $J_{dd}$ to be identical to the values given in Ref. 15, but rescaled by a factor of 0.36 to fit the experimentally observed critical temperature of $T_C = 398\,\mathrm{K}$. For comparison with our experiment, we finally determine the Faraday rotation resulting from the calculated sublattice magnetizations $M_l$ (see Eq. (1)). The results of this procedure are shown and discussed in Figs. 5A and S6.

**ACKNOWLEGMENT**

**General**: We thank G.E.W. Bauer and S.M. Rezende for stimulating discussions.

**Funding**:

T.K. acknowledges funding through the grants European Research Council H2020 CoG "TERAMAG" (grant no. 681917) and H2020 FET-OPEN project "ASPIN" (grant no. 766566) and DFG SFB/TRR 227 "Spin Dynamics" (project A05). I.R. acknowledges funding through German Federal Ministry of Education and Research (BMBF) grant 05K16BCA "Femto-THz-X." P.M. and P.M.O. acknowledge financial support from the Swedish Research Council (VR), the K. and A. Wallenberg Foundation (grant no. 2015.0060), and the Swedish National Infrastructure for Computing. M.G. thanks for funding through BMBF grant no. 05K10KEB. R.V.P. acknowledges financial support from the Russian Scientific Foundation (grant no. 16-12-10456). M.W. acknowledges funding by the Max Planck Society.

**Author contributions**: I.R. conceived the original experiment idea. S.F.M., I.R. and T.K. designed the experiment. S.F.M. built the setup, performed measurements and analyzed data. I.R., A.M.K. and R.V.P. prepared the samples, which were characterized by S.F.M., I.R., A.M.K., A.P. and M.G. The theoretical model was developed by T.K. and S.F.M., with contributions from J.B., P.M., A.P. and P.M.O. Atomistic spin-dynamics simulations were conducted by J.B. Equilibrium magnetization curves were calculated by S.F.M. and T.K. *Ab initio* calculations of electronic and vibrational properties were performed by P.M. and P.M.O. The manuscript was written by T.K., J.B. and S.F.M., with contributions from A.P., P.M.O., P.M., M.W. and I.R. All authors contributed to the discussions of the results and commented on the manuscript.

# FIGURES

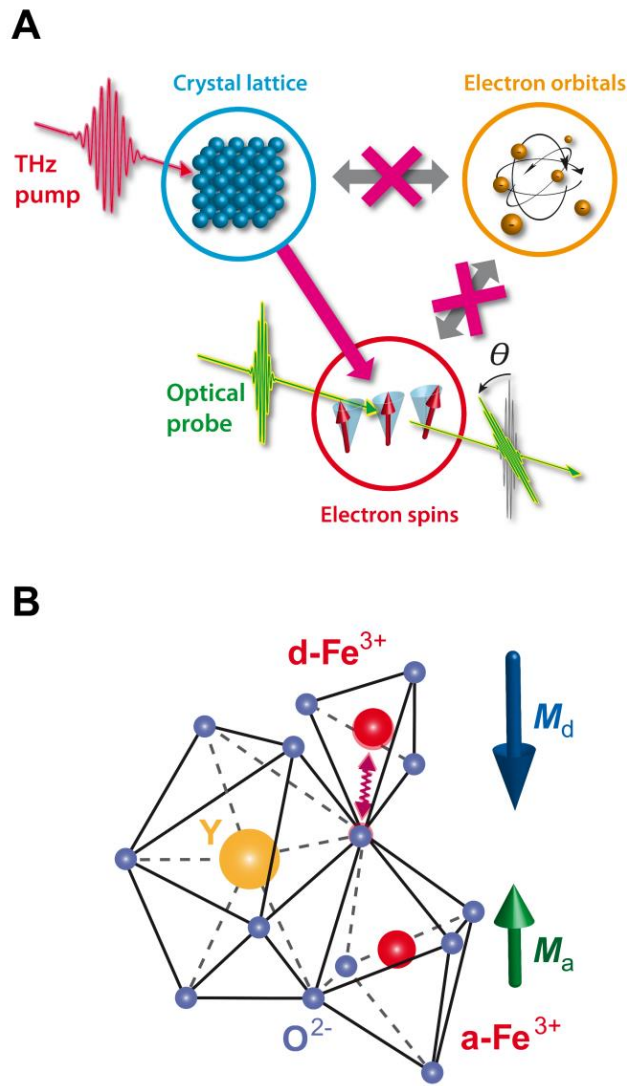

**Fig. 1. Ultrafast probing of spin-phonon interactions.** (**A**) Experimental principle. A THz pump pulse resonantly and exclusively excites optical phonons of a ferrimagnet. The impact on the sample magnetization is monitored by the Faraday rotation $\theta$ of a subsequent femtosecond probe pulse. By using an electric insulator, the electronic orbital degrees of freedom remain unexcited (see red crosses). (**B**) Part of the unit cell of ferrimagnetic YIG. Magnetic $Fe^{3+}$ ions at tetrahedral d-sites and octahedral a-sites comprise, respectively, the majority and minority spin-sublattice of the ferrimagnet. The pump pulse resonantly excites a TO($\Gamma$) optical phonon associated with a Fe-O stretch vibration at the tetrahedral d-site.

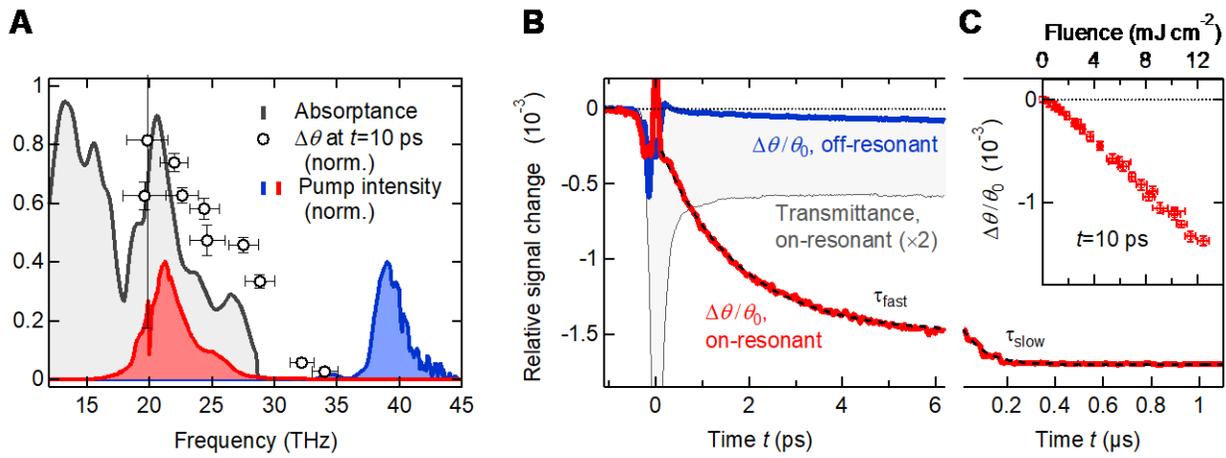

**Fig. 2. Ultrafast phonon-induced dynamics of magnetic order.** (**A**) THz absorptance of the BiGa:YIG film (black solid line). Pump intensity spectra are either resonant (red) or non-resonant (blue) with the TO($\Gamma$) phonon absorption band. Open circles show the pump-induced Faraday signal 10 ps after sample excitation as a function of the pump pulse center frequency. (**B**) Pump-induced change $\Delta\theta$ in Faraday rotation for resonant and off-resonant pumping on ultrafast and (**C**) microsecond time scales normalized to the equilibrium Faraday signal $\theta_0=\theta(-2\ \text{ps})$. The incident fluence is 10 mJ cm$^{-2}$. Panel (B) also shows the isotropic transient change in the sample transmittance for resonant pumping (thin black solid line). Dashed lines in panels (B) and (C) are single-exponential fits with time constants of $\tau_{\text{fast}}=1.6$ ps and $\tau_{\text{slow}}=90$ ns, respectively. The inset of panel (C) displays the ultrafast Faraday signal $\Delta\theta(10\ \text{ps})$ as a function of the incident pump-pulse fluence. Data are taken at a temperature of 296 K.

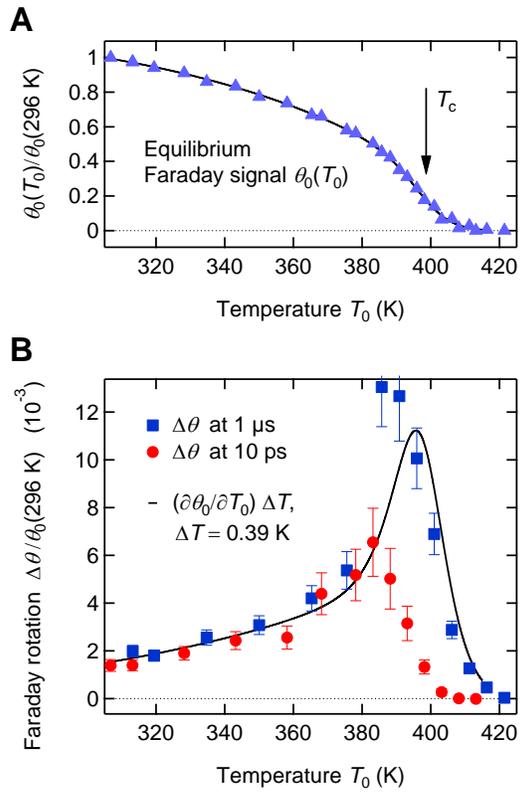

**Fig. 3. Two regimes of spin-lattice equilibration.** (**A**) Equilibrium Faraday rotation $\theta_0=\theta(-2\text{ ps})$ vs ambient temperature $T_0$ along with a fit to an analytical function (thin solid line). (**B**) Pump-induced change $\Delta\theta$ in the Faraday rotation at $t=10$ ps (red symbols) and 1 µs (blue) after pump-pulse arrival. The black curve is the change $(\partial\theta_0/\partial T_0)\Delta T$ in the Faraday rotation expected from the increase $\Delta T$ of the sample temperature due to heating by the pump pulse. $\theta_0(T_0)$ is taken from panel (A) (thin solid line), and $\Delta T=0.39$ K is calculated from the absorbed pump energy and the heat capacity of the excited volume.

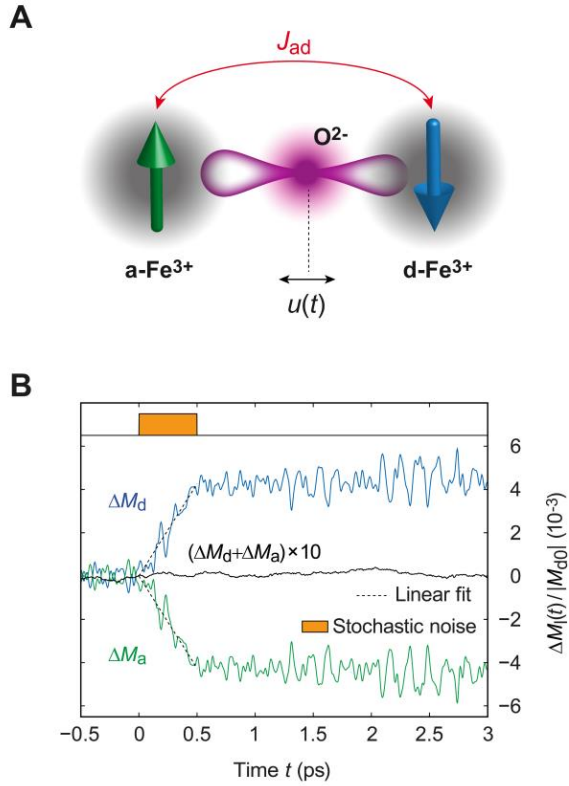

**Fig. 4. Atomistic spin-dynamics simulations.** (**A**) Schematic of our model of ultrafast spin-phonon coupling. Thermal motion of the $O^{2-}$ ion modulates the superexchange constant $J_{ad}$ of the adjacent a-$Fe^{3+}$ and d-$Fe^{3+}$ spins, thereby enabling transfer of spin angular momentum between a- and d-spin-sublattices. (**B**) Evolution of a- and d-spin-sublattice magnetizations as obtained by atomistic spin-dynamics simulations. From 0 to 0.5 ps, thermal modulation of the exchange constant $J_{ad}$ is switched on (orange square). The variance of the $J_{ad}$ fluctuation is proportional to the difference $\Delta T$ between crystal-lattice and spin temperature. To obtain agreement of the slope of $\Delta M_a(t)$ with that found in the experiment directly after pump excitation ($-0.1\%$ ps$^{-1}$, see Fig. 2B) where $\Delta T=0.39$ K (see Fig. 3B), an approximately three times smaller $\partial J_{ad}/\partial u$ of $\sim 10\ J_{ad}$ Å$^{-1}$ than used here has to be chosen.

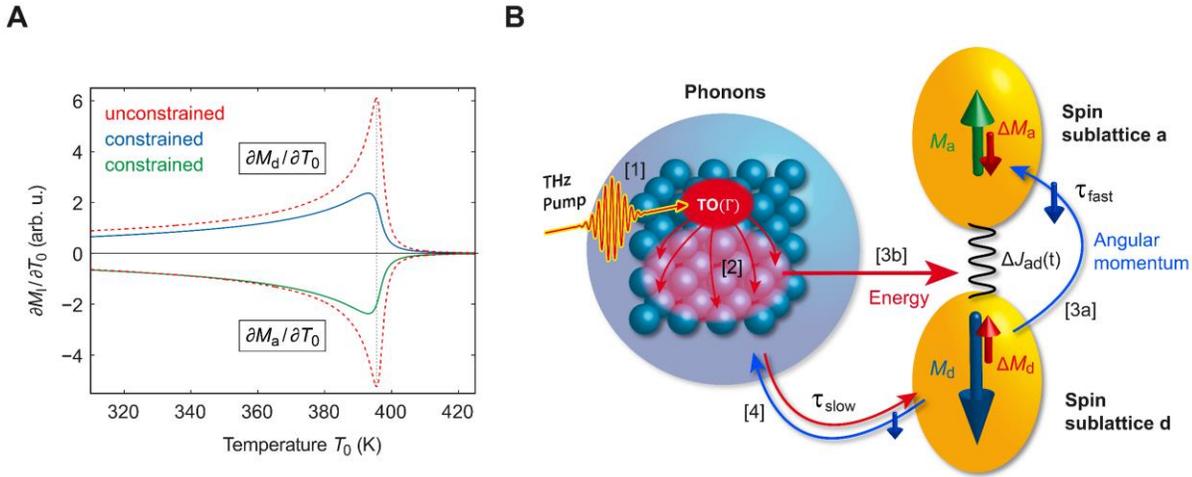

**Fig. 5. Constrained state and spin-phonon equilibration in YIG.** (**A**) Calculated change in sublattice magnetization $M_a$ and $M_d$ per increase of temperature $T_0$, without and with the constraint of constant spin angular momentum. The constrained and unconstrained $\partial M_d/\partial T_0$ curves exhibit good qualitative agreement with the measured Faraday signals $\Delta\theta(10\text{ ps})$ and $\Delta\theta(1\text{ µs})$ vs $T_0$ shown in Fig. 3B. (**B**) Schematic of spin-phonon equilibration in YIG. [1] Pump-excited TO($\Gamma$) phonons [2] increase the population of other lattice modes. The increased thermal modulation of the a-d-exchange by $\Delta J_{ad}(t)$ leads to [3a] transfer of angular momentum between a- and d-spin-sublattices, accompanied by [3b] energy transfer from the phonon to the spin system on the time scale $\tau_{fast}$=1.6 ps. The resulting state is constrained by $\Delta M_a+\Delta M_d=0$ and decays by [4] transfer of angular momentum and energy between crystal lattice and electron spins on the $\tau_{slow}$=90 ns scale. This process is mediated by spin-orbit and/or spin-spin magnetic-dipole coupling.

**Supplementary Material References: 46, 47, 48**